# A representative sampling plan for auditing health insurance claims[*]

## Arthur Cohen[1] and Joseph Naus[1]

*Rutgers University*

**Abstract:** A stratified sampling plan to audit health insurance claims is offered. The stratification is by dollar amount of the claim. The plan is representative in the sense that with high probability for each stratum, the difference in the average dollar amount of the claim in the sample and the average dollar amount in the population, is "small." Several notions of "small" are presented. The plan then yields a relatively small total sample size with the property that the overall average dollar amount in the sample is close to the average dollar amount in the population. Three different estimators and corresponding lower confidence bounds for over (under) payments are studied.

## 1. Introduction

Auditing health insurance claims is of necessity, extensively practiced. Statistical sampling plans and analysis of data from such plans is also extensive and diverse. One recent bibliography is Sampling for Financial and Internal Audits compiled by Yancey [5]. An annotated bibliography is given in Statistical Models and Analysis in Auditing (SMAA) [3] compiled by the Panel on Nonstandard Mixtures of Distributions, Committee on Applied and Theoretical Statistics of the Board on Mathematical Sciences, National Research Council. Most standard sampling methodologies have been considered. Namely, simple random sampling, stratified random sampling, and dollar unit sampling. In addition a variety of estimators of overpayments (underpayments) have been considered. (See SMAA [3].) These include the mean-per-unit estimator, the difference estimator, two types of ratio estimators, weighted averages of the above three, dollar unit estimator based on estimating the proportion of items in which overpayment occurs and Stringer estimators based on an estimator of the above proportion and also the data corresponding to actual overpayments (underpayments).

In this study we seek a stratified sampling plan whose main objective is to produce a lower confidence bound for the amount of overpayments (underpayments). Stratification is done on dollar amount of the claim (book). It is envisioned that this lower bound could justify a repayment by a health care provider to a client whose employees are covered by the provider's health insurance plan. Desirable features sought include simplicity, representativeness, relatively small total sample size yet somewhat adequate sample size in each stratum, no samples from zero dollar claims and relatively larger samples from strata with high dollar claims. Furthermore the

[*]Research supported by NSF Grant DMS-0457248 and NSA Grant H 98230-06-1-007.
[1]Department of Statistics and Biostatistics, Rutgers University, Hill Center, Busch Campus, 110 Frelinghuysen Road, Piscataway NJ 08854-8019, USA, e-mail: artcohen@rci.rutgers.edu; naus@rci.rutgers.edu

*AMS 2000 subject classifications:* primary 60K35; secondary 60K35.

*Keywords and phrases:* stratified sampling plan, dollar amount of claim, over (under) payments, unbiased estimator, separate ratio estimator, combined ratio estimator, lower confidence bound.

plan should audit all extremely high dollar claims and treat them separately. That is, such claims should not be included as part of the statistical sample.

To achieve the stated objectives we propose a "dollar representative stratified sampling plan" and refer to it as RepStrat sampling. The strata are class intervals of dollar amounts of claims. Class boundaries of strata are chosen according to typical guidelines (ample number of claims in each stratum, rounded numbers for boundaries, not too few, not too many). In addition the strata are adjustable so that the sample from each stratum is chosen in such a way that the average dollar amount of the claim for the population is "close" to the average dollar amount of the claim in the sample. The notion of "close" is made explicit in the next section. The closeness of the average dollar amounts of claims in sample and in population is what we mean by representativeness. Sample sizes for each stratum are chosen to ensure this closeness with high probability. We demonstrate that closeness within each stratum guarantees a higher degree of closeness between the averages in the overall population and overall sample.

Representativeness, as measured by closeness of average dollar amounts in sample and population, is important for several reasons. First, it is intuitively desirable. Second, oftentimes estimates of overpayments are needed by an agency or company in order to recover money that was excessively paid out. Since larger dollar amounts of claims have more opportunity for larger overpayments the agency would welcome higher average dollar amounts in the sample strata than in the population strata. On the other hand the agency's adversary would prefer smaller average dollar amounts in the sample. Since there is always a chance of litigation for the sake of recovering money, the least biased situation is to have representativeness in terms of the closeness notion.

Representativeness is not the only feature of the plan proposed here. We want the sample size to be moderate. Not too large because of auditing expense and yet large enough to get an estimate of overpayment whose variance is not too large. Furthermore choosing samples within strata randomly allows for plausible estimators of means and variances in an unbiased way. Still further the plan offered here is not model based and does not require distributional or other assumptions. Thus the plan offered here is a balance of a sense of fairness in a litigation setting. Representativeness, randomness, distributional robustness, and adequate sample size.

In Section 2 several different definitions of closeness will be offered and their properties will be studied. In Section 3 we will display customary formulas for lower confidence bounds based on several different estimators of overpayments. Section 4 contains an example. Standard textbook references containing sections on stratified sampling are Cochran [1], Scheaffer, Mendenhall, and Ott, 6th edition [4], and Lohr [2].

## 2. Determination of strata and sample sizes

Let $N$ be the total number of claims in the population under study. Each claim has a dollar amount (book) and the first step in a stratified sampling plan is to determine $L$ strata. The $L$ strata will be formed as class intervals $(A_i, B_i)$, $i = 1, 2, \ldots, L$. These $L$ strata can be determined iteratively if necessary so that certain properties of the sample (prior to auditing) are achieved. Justification for stratification is an intuitive impression that higher book values are correlated with higher or greater likelihood of overpayments. In our case representativeness is easier to achieve with stratification. In addition we want to assure that relatively more samples are drawn from higher dollar strata.

Once strata boundaries are determined (not necessarily finalized) sample sizes for each stratum need to be determined. In general, the researcher seeks to estimate a population parameter (say $\mu$) within a certain precision $g$, with a certain confidence level, $1 - \alpha$. In stratified sampling $n_i$ is the sample size for each stratum and the total sample size is $n = \sum_{i=1}^{L} n_i$. Let $Y_{ij}$ be the known dollar amount of the $j$th claim in the $i$th stratum, $i = 1, \ldots, L$; $j = 1, \ldots, N_i$, where $N_i$ is the number of claims in stratum $i$. Let $\bar{Y}_i = \sum_{i=1}^{N_i} Y_{ij}/N_i$ and $V_i = \sum (Y_{ij} - \bar{Y}_i)^2/N_i$ be the mean and variance respectively of the $Y_{ij}$ in stratum $i$. Let $y_{ij}$ and $x_{ij}$ be respectively the sample book amount and audited amount of the $j$th claim in the $i$th stratum, $i = 1, \ldots, L$; $j = 1, \ldots, n_i$. Also let $d_{ij} = \max(0, y_{ij} - x_{ij})$ be the sample amount overpaid on an audited claim. The $i$th stratum mean and variance of $d_{ij}$ are respectively denoted by $\mu_i$ and $\sigma_i^2$, and the mean of the overpayment variable for the population is denoted by $\mu$ where $\mu = \sum N_i \mu_i/N$. Finally let $W_i = N_i/N$ and $w_i = n_i/n$.

Note that when $x_{ij} \neq y_{ij}$ that means the auditor has detected an error in treating the claim. A desirable or even acceptable error rate (that includes underpayments as well as overpayments) is 1%, which is the standard in some industries. However it is not uncommon to see higher rates of 3 to 5 or even 8 percent. In the latter cases overpayments or even overpayments minus underpayments can be in the millions of dollars. In [3] the $d_{ij}$'s are modeled as coming from a mixture distribution. One distribution of the mixture is degenerate at the point $\{0\}$. More discussion regarding the other distribution is given in [3] where it is mentioned that the distribution may depend on the book amount. In our study no distributional assumptions are made.

There are a variety of ways sample sizes are determined or allocated to the strata. These include equal allocation ($w_i = 1/L$), proportional allocation ($w_i = W_i$) and Neyman allocation ($w_i = W_i \sigma_i / \sum W_i \sigma_i$). Neyman allocation is "optimal" in the sense of requiring the least total sample size to achieve the required overall precision. However knowledge of $\sigma_i$ is rarely available and proportional allocation is often preferred to Neyman allocation since it offers a type of "representativeness" in the sense that each stratum appears in the sample the same fraction of the time that it appears in the population. Researchers are sometimes willing to take a somewhat larger sample for "representativeness" and for the simplicity gained.

In RepStrat sampling we allocate the sample to gain representativeness in the dollar amount of the claims in each stratum and in the overall population. Toward this end the researcher specifies a level of precision $g_i$ and a confidence level $1 - \gamma$ such that the sample is representative for each stratum in the sense that

$$(2.1) \qquad P\{|\bar{y}_i - \bar{Y}_i| \leq g_i\} \geq 1 - \gamma, \qquad i = 1, \ldots, L.$$

The practitioner may specify equal absolute precision in estimating stratum means, by choosing equal values for $g_i$. This is considered in more detail under case (a) below. Alternatively, the practitioner may specify equal relative precision in estimating stratum means, i.e., $g_i = f\bar{Y}_i$, $0 < f < 1$; case (b) below deals with this case. Other choices for specifying stratum precision are considered in cases (c) through (e) below. We show that certain types of specifications are related to proportional allocation in stratified random sampling (case c), or Neyman optimal allocation (case d). We first deal with general $\{g_i\}$.

Given the practitioner specifies the desired stratum precisions in terms of the $\{g_i\}$ and the confidence $1 - \gamma$, then the stratum sizes $\{n_i\}$ can be determined as follows. Strata boundaries are chosen so that the sample sizes determined will hopefully be adequate enough so that $\bar{y}_i$ will be approximately normal. In light of this and (2.1) we find $n_i$ so that

$$(2.2) \qquad P\left\{|Z| < g_i \sqrt{\frac{n_i}{V_i} \frac{N_i - 1}{(N_i - n_i)}}\right\} \geq 1 - \gamma,$$

where $Z$ is a standard normal variable. This leads to

$$(2.3) \qquad n_i = z_{\gamma/2}^2 V_i N_i \big/ \big[g_i^2(N_i - 1) + z_{\gamma/2}^2 V_i\big],$$

where $z_{\gamma/2}$ is the $1 - \gamma/2$ percentile of a standard normal.

Should $N_i$ be large so that the finite population correction factor (fpc) be close to 1, then (2.3) reduces to

$$(2.4) \qquad n_i = z_{\gamma/2}^2 V_i / g_i^2.$$

**Remark 2.1.** Typically strata with larger dollar amounts of claims will also have larger values of $V_i$. Thus from (2.3) we see that relatively larger samples will be drawn from such strata. This was one of the properties felt to be desirable in a sampling plan of this type.

Now let $\bar{y}_{st} = \sum N_i \bar{y}_i / N$ and $\bar{Y} = \sum N_i \bar{Y}_i / N$ be respectively the sample estimate of and true population mean of book dollar amounts. Given the stratum $\{n_i\}$ are determined from the stratum precisions $\{g_i\}$ and $1-\gamma$, and given the known stratum variances $V_i$ of the $Y_{ij}$'s yields the variance of $\bar{y}_{st}$. We can use this to approximate the distribution of $|\bar{y}_{st} - \bar{Y}|$. For a specified value of $g$, we can approximate $P\{|\bar{y}_{st} - \bar{Y}| \leq g\}$; we will denote this probability by $1 - \alpha$. For a chosen value of $g$ we consider

$$(2.5) \quad P\{|\bar{y}_{st} - \bar{Y}| \leq g\} = P\left\{|\bar{y}_{st} - \bar{Y}| \big/ \sqrt{\sum W_i^2 V_i / n_i} \leq g \big/ \sqrt{\sum W_i^2 V_i / n_i}\right\}.$$

If we ignore the fpc and substitute (2.4) in (2.5), we find that (2.5) is approximately

$$(2.6) \qquad P\left\{|Z| \leq z_{\gamma/2} \big/ \sqrt{\sum_{i=1}^{L} W_i^2 (g_i/g)^2}\right\}.$$

Thus if

$$(2.7) \qquad \sum_{i=1}^{L} W_i^2 (g_i/g)^2 \leq 1$$

it follows from (2.6) that the probability that the overall sample mean of claims is close to the population mean of claims is at least $1 - \gamma$. That is, if (2.7) holds

$$(2.8) \qquad P\{|\bar{y}_{st} - \bar{Y}| \leq g\} = 1 - \alpha \geq 1 - \gamma.$$

Clearly in this instance $\alpha \leq \gamma$. Note that (2.7) and (2.8) are easily satisfied in the case where $g_i = g$, for all $i$; they are also satisfied if $g_i = f\bar{Y}_i$ and $g = f\bar{Y}$ for constant $f$. Thus, in the case where the stratum mean estimates precisions are equal (either absolutely or relative to the means), the combined stratum estimate has at least as good precision.

In the auditing application considered in this paper, claims are stratified by claim amount $Y_i$ into $L$ strata. A special type of stratified random sample is picked with $n_i$ claims from the $N_i$ claims within the $i$th stratum, for $i = 1, \ldots, L$. The stratum sample sizes are chosen to make the sample "representative" in that for every stratum, there is a high probability $1 - \gamma$ that the stratum sample mean will be "close" (for the $i$th stratum, within $g_i$) of the true stratum mean. Given the $\{g_i\}$ and $\gamma$, the $\{n_i\}$ are completely determined, as is $P\{|\bar{y}_{st} - \bar{Y}| \leq g\}$, for any $g$. The formula (2.1) through (2.8) detail this.

In general stratified random sampling the desired closeness of $\bar{y}_{st}$ to $\bar{Y}$ is specified together with some type of allocation to determine the $\{n_i\}$. This in turn determines the within stratum precision which is typically of minor or secondary interest. By contrast, the representative stratified random sampling approach specifies the desired "closeness" (of $\bar{y}_i$ to $\bar{Y}_i$) within individual stratum to find the $\{n_i\}$. In the next section we relate these two approaches to specifying sample sizes in stratified random sampling.

## 3. Relation between representative and general stratified sampling approaches

In general stratified random sampling, some particular method of allocating the total sample size $n$ to the $L$ individual strata is chosen. In equal allocation, $n_i = n/L$. In proportional allocation, $n_i = nN_i / \sum N_i$; that is the $n_i$ are proportional to the $N_i$.

How does representative stratified random sampling relate to various other types of allocation in stratified random sampling? To analyze this, fix the precision of the estimator $\bar{y}_{st.}$; that is specify $g$ and $\alpha$ in equation (2.8). Various ways to specify the within stratum precision (the $g_i$'s and $\gamma$) are related to types of allocation (proportional, optimal and other) in stratified random sampling. For a given desired precision of the estimator $\bar{y}_{st}$ given by $\alpha$ and $g$, and choice of spcification of $g_i$, and $\gamma$, we can find the total sample size $n$, as well as the sample weights $w_i = n_i/n$. For example, we will show that if the $g_i$ are all taken to be equal, then $w_i$ is proportional to the stratum variances, and the overall $n$ is given by equation (3.4) below. More generally, we show how to determine $\{n_i\}$ and $n$ given any three of $\{g_i\}$, $\alpha$, $\gamma$, $g$.

Divide both the $|\bar{y}_{st} - \bar{Y}|$ and $g$ in (2.8) by $\sqrt{\sum_{i=1}^{L} W_i^2 V_i/n_i}$; use the normal approximation and ignore the fpc, to find that approximately

$$\sum_{i=1}^{L} W_i^2 V_i/n_i = g^2/z_{\alpha/2}^2.$$

Thus choosing $n_i$ as in (2.4) yields

$$(3.1) \qquad \sum W_i^2 g_i^2 = g^2 z_{\gamma/2}^2 / z_{\alpha/2}^2.$$

At this point we present a variety of choices of $g_i$.

Case (a): $g_i = C$ for all $i = 1, \ldots, L$.
Then from (3.1)

$$(3.2) \qquad C^2 = g^2 z_{\gamma/2}^2 / z_{\alpha/2}^2 \sum_{i=1}^{L} W_i^2,$$

and should we want $n_i$ to satisfy both (2.1) and (2.8), we have from (2.4) that

$$(3.3) \qquad n_i = V_i z_{\alpha/2}^2 \sum_{i=1}^{L} W_i^2 / g^2.$$

That is, $n_i$ is proportional to $V_i$. The overall sample size in this case is

$$(3.4) \qquad n = (z_{\alpha/2}/g)^2 \sum_{i=1}^{L} W_i^2 \sum_{j=1}^{L} V_j$$

and the sample weights $w_i = n_i/n$ are given by

$$(3.5) \qquad w_i = V_i \Big/ \sum_{i=1}^{L} V_i.$$

Case (b): $g_i = f\bar{Y}_i$ for all $i = 1, \ldots, L$.
If $g_i$ is a fixed proportion of strata mean book amount, then from (3.1) we have approximately

$$(3.6) \qquad f^2 = g^2 z_{\gamma/2}^2 / z_{\alpha/2}^2 \left( \sum_{i=1}^{L} W_i^2 \bar{Y}_i^2 \right)$$

and thus if we want $n_i$ to satisfy both (2.1) and (2.8), we have from (2.4)

$$(3.7) \qquad n_i = z_{\alpha/2}^2 V_i \left( \sum_{j=1}^{L} W_j^2 \bar{Y}_j^2 \right) \Big/ g^2 \bar{Y}_i^2.$$

Thus in this case $n_i$ is proportional to $V_i/\bar{Y}_i^2$, which is the squared coefficient of variation of the book value for the $i$th situation. The sample weights $w_i = n_i/n$ are

$$(3.8) \qquad V_i/\bar{Y}_i^2 \left( \sum_{i=1}^{L} V_i/\bar{Y}_i^2 \right).$$

Case (c): $g_i = f\sqrt{V_i/W_i}$.
From (2.4)

$$(3.9) \qquad n_i = z_{\gamma/2}^2 W_i / f^2 = K W_i,$$

when $K$ is constant. This amounts to proportional allocation related to the number of claims in a stratum. This method of sample size determination does not really satisfy our goals.
Case (d): $g_i = f V_i^{1/4}/\sqrt{W_i}$.
From (2.4)

$$(3.10) \qquad n_i = z_{\gamma/2}^2 W_i \sqrt{V_i} / f^2.$$

From (3.1)

$$(3.11) \qquad f = g z_{\gamma/2}/z_{\alpha/2} \sqrt{\sum W_i V_i^{1/2}}.$$

This is Neyman-optimum allocation for estimating mean claim size. For this case using (3.10) and (3.11)

$$(3.12) \qquad n = \left( \sum_{i=1}^{L} W_i V_i^{1/2} \right)^2 z_{\alpha/2}^2 / g^2$$

and

$$(3.13) \qquad w_i = W_i V_i^{1/2} \bigg/ \sum_{i=1}^{L} W_i V_i^{1/2}.$$

Case (e): $g_i = f \bar{Y}_i^{1/2}$.

This designation of strata precision is a compromise between case (a) which seeks the same absolute precision for each stratum and case (b) which seeks the same relative precision. Here, using (2.4) and (3.1)

$$f = g z_{\gamma/2}/z_{\alpha/2} \sqrt{\sum W_i^2 \bar{Y}_i},$$

$$(3.14) \qquad n_i = V_i z_{\alpha/2}^2 \sum_{j=1}^{L} W_j^2 \bar{Y}_j / \bar{Y}_i g^2$$

so

$$(3.15) \qquad w_i = (V_i/\bar{Y}_i) \bigg/ \sum_{j=1}^{L} V_j/\bar{Y}_j.$$

**Remark 3.1.** In deriving most formulas for $n_i$ we have been ignoring the fpc. Should the fpc not be close to 1 a modification of the formulas may be necessary. The modification entails replacing $V_i$ by $V_i(N_i - n_i)/(N_i - 1)$ and solving the resulting equation for $n_i$. So for example in case (a), (3.3) would be replaced by

$$(3.16) \qquad n_i = N_i z_{\alpha/2}^2 \sum_{j=1}^{L} W_j^2 \bigg/ \left[ N_i - 1 + V_i z_{\alpha/2}^2 \sum_{j=1}^{L} W_j^2 \right].$$

In case (b) the $n_i$ of (3.7) would be replaced by

$$(3.17) \qquad n_i = \left[ V_i z_{\alpha/2}^2 N_i \left( \sum_{j=1}^{L} W_j^2 \bar{Y}_j^2 \right) \right] \bigg/ \left[ (N_i - 1) g^2 \bar{Y}_i^2 + V_i z_{\alpha/2}^2 \sum_{j=1}^{L} W_j^2 \bar{Y}_j^2 \right].$$

Of course if we were concerned only with $n_i$ satisfying (2.1), recognizing the implications this has for (2.8), then $n_i$ can be determined from (2.3) which includes the fpc.

## 4. Estimating overpayments

In this section we describe several point estimators and corresponding lower confidence bounds for $(OP)$, the total amount overpaid. Most of the material here is essentially drawn from the cited SMAA [3] article appearing in *Statistical Science*, and Cochran [1]. We focus on three of the estimators although the SMAA article discusses others as well. It would not be an unreasonable approach, when seeking recovery of money, to give several estimates of $(OP)$ to see if they agree. If they do not, the parties might choose to compromise among the competing estimators.

We are somewhat guided in the selection of the three estimators by the findings of the SMAA committee in the sense that some of the following estimators lead to conservative lower bounds. The three estimators are

**The difference estimator:**

$$(4.1) \qquad (\widehat{OP})_d = \sum N_i \bar{d}_i,$$

where

$$\bar{d}_i = \sum_{j=1}^{n_i} d_{ij}/n_i.$$

**The separate ratio estimator:**

$$(4.2) \qquad (\widehat{OP})_{RS} = \sum_{i=1}^{L} N_i \bar{Y}_i r_i,$$

where

$$r_i = \bar{d}_i/\bar{y}_i.$$

**The combined ratio estimator:**

$$(4.3) \qquad (\widehat{OP})_{RC} = r_c \sum_{i=1}^{L} N_i \bar{Y}_i,$$

where

$$(4.4) \qquad r_c = \sum_{i=1}^{L} N_i \bar{d}_i \Big/ \sum_{i=1}^{L} N_i \bar{y}_i.$$

Denote the sample variance of $d_{ij}$ by $s_{d_i}^2$, where

$$(4.5) \qquad s_{d_i}^2 = \sum_{j=1}^{n_i} (d_{ij} - \bar{d}_i)^2/(n_i - 1).$$

Then the estimated variance of $(\widehat{OP})_d$ is

$$(4.6) \qquad \hat{V}(\widehat{OP})_d = \sum_{i=1}^{L} N_i(N_i - n_i)s_{d_i}^2/n_i.$$

A $(1 - \beta)$ lower confidence bound based on an estimate $\hat{\theta}$ for total overpayment is

$$(4.7) \qquad \hat{\theta} - z_\beta \big(\hat{V}(\hat{\theta})\big)^{1/2}.$$

Note that $\hat{V}(\widehat{OP})_d$ depends on the $s_{d_i}^2$ which can be computed from (4.5), or in the case where many of the $d_i$'s are zero, more simply from

$$s_{d_i}^2 = \left\{ n_i \sum_{j=1}^{n_i} d_{ij}^2 - \left( \sum_{j=1}^{n_i} d_{ij} \right) \left( \sum_{j=1}^{n_i} d_{ij} \right) \right\} \Big/ n_i(n_i - 1),$$

where

$$\sum_{j=1}^{n_i} d_{ij} \qquad \text{and} \qquad \sum_{j=1}^{n_i} d_{ij}^2$$

can be computed from just the non-zero $d_{ij}$. This type of simplification is of even greater help for the ratio estimates that follow.

The estimated variance of $(\widehat{OP})_{RS}$ is

$$(4.8) \qquad \hat{V}(\widehat{OP})_{RS} = \sum_{i=1}^{L} N_i s_{RSi}^2 (N_i - n_i)/n_i,$$

where

$$\begin{aligned}
s_{RSi}^2 &= \sum_{j=1}^{n_i} (d_{ij} - r_i y_{ij})^2/(n_i - 1) \\
(4.9) \qquad &= s_{d_i}^2 + r_i^2 s_{y_i}^2 - 2r_i s_{d_i y_i}
\end{aligned}$$

and

$$\begin{aligned}
r_i &= \bar{d}_i/\bar{y}_i, \\
s_{d_i y_i}^2 &= \sum_{j=1}^{n_i} (d_i - \bar{d}_i)(y_i - \bar{y}_i)/(n_i - 1) \\
&= \left\{ \sum_{j=1}^{n_i} d_{ij} y_{ij} - \bar{y}_i \sum_{j=1}^{n_i} d_{ij} \right\} \Big/ (n_i - 1).
\end{aligned}$$

Note that $s_{d_i y_i}^2$ and $s_{d_i}^2$ can be computed from $\sum d_{ij}$, $\sum d_{ij}^2$, and $\sum d_{ij} y_{ij}$, which can be computed from just the non-zero $d_{ij}$. Thus, just knowing $(d_{ij}, y_{ij})$ for the non-zero $d_{ij}$, and $\bar{y}_i$ and $s_{y_i}^2$ are sufficient to compute (4.8).

The estimated variance of $(\widehat{OP})_{RC}$ is

$$(4.10) \qquad \hat{V}(\widehat{OP})_{RC} = \sum_{i=1}^{L} N_i s_{ri}^2 (N_i - n_i)/n_i,$$

where

$$\begin{aligned}
s_{ri}^2 &= \sum_{j=1}^{n_i} (d_{ij} - r_c y_{ij})^2/(n_i - 1) \\
(4.11) \qquad &= \left\{ \sum_{j=1}^{n_i} d_{ij}^2 + r_c^2 \sum_{j=1}^{n_i} y_{ij}^2 - 2r_c \sum_{j=1}^{n_i} d_{ij} y_{ij} \right\} \Big/ (n_i - 1)
\end{aligned}$$

and $r_c$ is defined in (4.4).

Knowing $\bar{y}_i$ and $s_{y_i}^2$ gives $\sum_{j=1}^{n_i} y_{ij}^2$, and knowing these together with the $(d_{ij}, y_{ij})$ for non-zero $d_{ij}$ is sufficient to compute (4.10).

## 5. Example

We have constructed the following fictitious data set to be similar in number of claims, and book and audited amount of claims in data sets we have seen. Table 1 contains population data broken down by class intervals (strata) of designated dollar amount of claims. For each stratum we give the number of claims, the population mean ($\bar{Y}_i$) and variance ($V_i$) of dollar amount of claim, and the sample sizes ($n_i$) determined from (2.3) for $\gamma = .05$, and $g_i = .05\bar{Y}_i$. Table 2 contains sample data for those pairs ($d_{ij}, y_{ij}$) with positive overpayment, $d_{ij} > 0$. Note that there are several sample pairs ($d_{ij}, y_{ij}$) where $y_{ij} - d_{ij} = x_{ij} = 0$. Such cases result when payments were made on claims that shouldn't have been paid. Table 3 contains the stratum sample means and variances for the $d_{ij}$ and $y_{ij}$.

Table 4 compares the difference estimate (4.1), the separate ratio estimator (4.2), and the combined ratio estimate (4.4) for the total amount of overpayment. Also included are the 95% lower confidence bounds for total overpayment based on each estimator derived using (4.7). First we note that $|\bar{y}_{st} - \bar{Y}| = |418.7500 - 417.9375| = 0.812475 < 0.02\bar{Y}$. This indicates that RepStrat has accomplished one of its main goals. Namely that the average dollar amount of the claims in the sample is very close to the average dollar amount of claims in the population.

From Table 4 we note that all three estimation procedures yield results that are very close to each other. In light of the SMAA report on conservativeness of the lower confidence bounds and the closeness of the three estimation procedures we feel that RepStrat has provided a satisfactory analysis of the data set.

Table 1
*Population data and sample sizes*

| Strata Number | Dollar Value Strata | Number of Claims | Population Mean | Population Variance | |
|---|---|---|---|---|---|
| | $N_i$ | $\bar{Y}_i$ | $V_i$ | $n_i$ | |
| (1) | 0–199 | 4000 | 120 | 703 | 74 |
| (2) | 200–499 | 2200 | 313 | 3,500 | 54 |
| (3) | 500–999 | 1000 | 620 | 10,000 | 39 |
| (4) | 1,000–1,999 | 500 | 1148 | 30,000 | 33 |
| (5) | 2,000–3,999 | 200 | 2374 | 110,000 | 27 |
| (6) | 4,000–6,999 | 100 | 5061 | 250,000 | 14 |
| | | 8000 | | | 241 |

Table 2
*Sample Pairs: $(d_{ij}, y_{ij})$ when $d_{ij} > 0$*

| Strata | |
|---|---|
| 1 | (9, 44), (105, 105), (57, 57), (143, 143) |
| 2 | (8, 288), (422, 422), (115, 380), (93, 455), (495, 495), (359, 359) |
| 3 | (530, 530), (76, 516), (12, 736), (124, 540), (54, 711), (96, 674) |
| 4 | (804, 1804), (628, 1000), (718, 1000), (475, 1000), (500, 1500), (800, 1500) |
| 5 | (1120, 2520), (2607, 2607), (389, 3456), (1990, 3265), (3900, 3900), (100, 3900), (1550, 3000), (500, 3000) |
| 6 | (1220, 6102), (1750, 6999), (3, 5232), (6900, 6900), (100, 6671), (1220, 6102) |
| height | |

Table 3
*Sample Statistics*

| Strata | $n_i$ | $\bar{y}_i$ | $d_i$ | $s_{y_i}^2$ |
|---|---|---|---|---|
| 1 | 74 | 115 | 4.2432 | 680 |
| 2 | 54 | 300 | 27.6296 | 3400 |
| 3 | 39 | 650 | 22.8718 | 10500 |
| 4 | 33 | 1200 | 118.9394 | 30300 |
| 5 | 27 | 2400 | 450.2222 | 111000 |
| 6 | 14 | 5000 | 799.5000 | 250000 |

Table 4
*Estimators and Lower 95% Confidence Bound*

| Type of Estimator | Estimator | Lower Confidence Bound |
|---|---|---|
| Difference | 330,094 | 214,037 |
| Separate Ratio | 329,833 | 220,286 |
| Combined Ratio | 329,453 | 215,323 |